\documentstyle[sprocl]{article}
\def\beq{\begin{equation}}
\def\eeq{\end{equation}}
\newcommand{\be}{\begin{equation}}
\newcommand{\ee}{\end{equation}}
\def\bea{\begin{eqnarray}}
\def\eea{\end{eqnarray}}

\def\Tr{\rm Tr}
\newcommand{\barre}[1]{%
	\setbox1=\hbox{$#1$} \dimen2=\ht1 \dimen3=\dp1 \dimen4=\wd1
	\setbox2=\hbox{\sl /}
	\dimen1=\wd1 \advance\dimen1 by -\wd2 \divide\dimen1 by 2
	\advance\dimen1 by \wd2 \advance\dimen1 by 0.4pt
	\setbox3=\hbox to \wd1{\hss \box1 \kern -\dimen1 \box2\hss}
	\ht3=\dimen2 \dp3=\dimen3 \wd3=\dimen4
	\box3}
\newcommand{\vev}[1]{%
	\langle #1 \rangle
	}

\def\1{{\rm 1 \kern -.10cm I \kern .14cm}} \def\R{{\rm R \kern -.28cm I
\kern .19cm}}
\begin{document}
\begin{flushright}    UFIFT-HEP-98-20 \\ 
%hep-ph/9808489 
\end{flushright}
\vskip .5cm
\title{A Model for Fermion Mass Hierarchies and Mixings}
\author{P. Ramond}
\address{Institute for Fundamental Theory, Physics Department.
\\ University of Florida, Gainesville, FL 32611} 
\vskip .5cm
\centerline{\it Invited Talk at PASCOS 98, May 1998, Boston, MA}
\vskip .5cm

\maketitle\abstracts{We present a model which contains three Abelian symmetries beyond the standard model. One is anomalous {\it \` a la} Green-Schwarz, and family-independent; the other two are family symmetries. All are broken at a large scale by string
y effects. The model is predictive in the neutrino sector: large mixing between $\nu_\mu-\nu_\tau$, and {\cal O}$(\lambda^3)$ $\nu_e-\nu_\mu$ mixings. Its natural cut-off is the gauge unification scale, orders of magnitude below the perturbative string cu
t-off. }  

\section{Introduction}
Simplicity in the minimal supersymmetric standard model appears only at the scale at which its gauge couplings unify, $M_{U}$, and where its Yukawa couplings  display {\bf two} different  hierarchies: interfamily hierarchy which relates and mixings partic
les of different families in each charge and color sector, as well as intrafamily hierarchy which relates particles of the same family. They have different theoretical origins. The former is due to non-anomalous family symmetries, the latter to family-ind
ependent anomalous symmetries. 

Our general framework is that of a low-energy effective theory with cut-off,  $M_{U}$. Anomalous gauge symmetries can exist in this framework as long as there is at cut-off an interaction which compensates for the Noether anomalies induced in the Lagrangi
an. String theories are such an example. The Green-Schwarz mechanism~\cite{GS} provides a  dimension-five interaction term, whose structure demands a specific pattern among the anomaly coefficients, namely that the combinations 
\be
\alpha_{\rm color}C_{\rm color}=
\alpha_{\rm weak}C_{\rm weak}=
\alpha_{\rm Y}C_{\rm Y}=\cdots=
\alpha_{\rm i}C_{\rm i}\ ,\ee
be universal, where $\alpha_i$ is the coupling constant of the gauge group $i$. Here 
\be
C_{\rm i}=\Tr(XG_iG_i)\ ,\ee
are the anomaly coefficients which appear in the divergence of the anomalous $X$-current. At $M_{U}$, these imply for the standard model
\be
\alpha_{\rm color}=\alpha_{\rm weak}~~\rightarrow~~ C_{\rm color}=C_{\rm weak}\ ,\ee
as well as
\be
\tan^2\theta_w={\alpha_{\rm weak}\over \alpha_{\rm Y}}={C_{\rm Y}\over C_{\rm weak}}\ .\ee
This relation relates an ultraviolet parameter, the Weinberg angle~\cite{Ib}, to infrared properties, since the anomaly coefficients are determined from the couplings of the massless particles in the theory. 

String theories with an anomalous Green-Schwarz $U(1)$, generate a Fayet-Iliopoulos $D$-term, which triggers the breaking of the anomalous symmetry and in general others at a scale that is computably lower than the cut-off. These models therefore generate
 a small expansion parameter, the ratio of the scale at which the anomalous symmetry is broken to the cut-off. 

The phenomenological requirement that neither supersymmetry nor standard model gauge symmetries be broken at a high scale severely restricts any theory that contains an anomalous $U(1)$~\cite{IL}. In particular, one can relate
the absence of dangerous flat directions to the presence of certain interaction terms in the superpotential. Remarkably, these are compatible with the invariants of the so-called minimal supersymmetric standard model. For example, the seesaw~\cite{SEESAW}
 mechanism was shown to imply~\cite{BILR} the absence of $R$-parity violating interactions.  

Moreover, the anomalous $U(1)$ offers a natural mechanism for supersymmetry breaking~\cite{BD}, once one assumes dilaton stabilization, and a non-Abelian gauge interaction other than QCD. 
\section{The Framework}
We consider models which have a gauge structure broken in two sectors: 
a visible sector, and a hidden sector, linked by the anomalous 
symmetry and possibly other Abelian symmetries (as well as gravity). 
\beq G_{\rm SM}\times U(1)_X\times U(1)_{Y^{(1)}}\cdots\times  
U(1)_{Y^{(M)}}\times G_{\rm hidden}\ ,\eeq  
where $G_{\rm hidden}$ is the hidden gauge group, and 
$ G_{\rm SM}$ is the standard model gauge group. Only $X$, 
is anomalous in the sense of Green-Schwarz.  $X$, $Y^{(a)}$ are spontaneously broken at a high 
scale by the Fayet-Iliopoulos term generated by the dilaton vacuum. This 
DSW vacuum~\cite{DSW} is required by phenomenology to preserve both supersymmetry and the standard model symmetries. 

The application of the Green-Schwarz structure to 
the standard model is consistent with many of its phenomenological 
patterns.
There are also intricate anomaly requirements as not all anomalies are accounted for by the Green-Schwarz mechanism. The anomalies are of the following type:
\begin{itemize}
\item  The first involve only standard-model gauge groups $G_{\rm SM}$, 
with coefficients $(G_{\rm SM}G_{\rm SM}G_{\rm SM})$, which cancel for 
each chiral family and for vector-like matter. Also the hypercharge mixed 
gravitational anomaly $(YTT)$ vanishes.
\item The second type is where the new symmetries 
appear linearly, of the type $(Y^{(i)}G_{\rm SM}G_{\rm SM})$. If we 
assume that the 
$Y^{(i)}$ are traceless over the three chiral families, these vanish over the three families of  fermions with standard-model charges. Hence they must vanish 
on the Higgs fields: with  $G_{\rm SM}=SU(2)$, it implies the Higgs pair is 
vector-like  with respect to the $Y^{(i)}$. It also follows that the mixed 
gravitational anomalies $(Y^{(i)}TT)$ are zero over the fields with 
standard model quantum numbers. 
\item The third type involve the new symmetries quadratically, of the form $(G_{\rm 
SM}Y^{(i)}Y^{(j)})$. These vanish by group theory except for those 
 of the form $(YY^{(i)}Y^{(j)})$. In general two types of 
fermions contribute: the three chiral families and 
standard-model vector-like pairs. 
\item The remaining vanishing anomalies involve the anomalous charge $X$. 
\end{itemize}
\begin{itemize} 
\item With $X$ family-independent, and 
$Y^{(i)}$ family-traceless,  the vanishing of the 
$(XYY^{(i)})$ anomaly coefficients   over the three families is
assured:  so they must also vanish over the Higgs 
pair. This means that $X$ is also vector-like on the Higgs pair. Hence  
 the standard-model invariant $H_uH_d$ (the $\mu$ 
term) has zero $X$ and $Y^{(i)}$ charges. In 
string theory, mass terms do not appear in the superpotential, but only 
in the K\"ahler potential. After 
supersymmetry-breaking, this generates an effective 
$\mu$-term, of weak strength, as suggested by Giudice and 
Masiero~\cite{GM}. 
\item The  coefficients $(XY^{(i)}Y^{(j)})$, $i\ne j$. Since 
standard-model singlets can contribute to these anomalies, we expect 
cancellation to 
come about through a combination of hidden sector and singlet fields.
\item The coefficient $(XXY)$. This imposes an important constraint on 
the $X$ charges on the chiral families.  
\item The coefficients $(XXY^{(i)})$; with family-traceless symmetries, 
they  vanish over the three families of 
fermions with standard-model charges, but contributions are expected from 
other sectors of the theory.
\end{itemize}
In the standard model, we have the three anomalies associated with its 
three gauge groups, 
\be 
C_{\rm color}=(XSU(3)SU(3))\ ;\ C_{\rm weak}=(XSU(2)SU(2))\ ;\ C_{\rm Y}=(XYY)\ ,\ee
where $()$ stands for the trace. They can be expressed~\cite{PMR}
in terms of the $X$-charges of the invariants of the MSSM
\be
C^{}_{\rm color}={1\over 2}\sum_{i}\left[X^{[u]}_{ii}+X^{[d]}_{ii}\right]
-3X^{[\mu]}_{}\ ,\ee
\be
C_{\rm Y}+C_{\rm weak}-{8\over 3}C_{\rm 
color}=2\sum_i\left[X^{[e]}_{ii}-X^{[d]}_{ii}\right]
+2X^{[\mu]}_{}\ ,\ee
where $X^{[u,d,e]}_{ij}$ are the $X$-charges of ${\bf Q}^{}_i{\bf\overline 
u}^{}_jH^{}_u$,  ${\bf Q}^{}_i{\bf\overline 
d}^{}_jH^{}_d$,  $L^{}_i{\overline e}^{}_jH^{}_d$ respectively, and finally
$X^{[\mu]}$ that of the $\mu$-term $H^{}_uH^{}_d$, 
where $i,j$ are the family indices. A top quark Yukawa mass coupling at 
tree-level, we have $X^{[u]}_{33}=X^{[u]}_{}=0$. This implies that the 
$X$-charge of the down quark Yukawa is proportional to the color anomaly, 
and  thus cannot vanish: the down Yukawa is  necessarily smaller 
than the top Yukawa, leading to the suppression of $m_b$ over $m_t$, 
after electroweak breaking! {\it The presence of the color anomaly 
implies suppression of the bottom mass relative to the top 
mass}. With $C_{\rm color}=C_{\rm weak}$, the second anomaly equation becomes  
\be
C_{\rm Y}-{5\over 3}C_{\rm weak}=6\left[X^{[e]}_{}-X^{[d]}_{}\right]
\ ,\ee
stating that the relative suppression of the 
down to the charged lepton sector is proportional to the difference of 
two anomaly coefficients. Since $m_b=m_\tau$ near $M_U$, this implies that 
\be
{3\over 5}={C_{\rm weak}\over C_{\rm Y}}=\tan^2\theta_w\ .\ee
This happens exactly at the phenomenologically 
preferred value of the Weinberg angle: the $b-\tau$ unification is
related to the value of the Weinberg angle~\cite{BR}!
\section{A Three-Family Model}
We can see how some of the features we have just discussed lead to
phenomenological consequences in the context of 
a three-family model~\cite{EIR,ILR}, 
with three Abelian symmetries broken in the DSW vacuum. The matter
content of the theory is inspired by $E_6$, which 
 contains two Abelian symmetries outside of the 
standard model: the two $U(1)$, $V'$, $V$, appear in the embeddings 
\be E_6\subset ~SO(10)\times U(1)_{V'}\qquad SO(10)\subset SU(5)\times U(1)_V\ .\ee
Over the three chiral families, the two non-anomalous symmetries are  
\beq Y^{(1)}={1\over 5}(2Y+V)
(2,-1,-1)\eeq
\beq Y^{(2)}={1\over 4}(V+3V')(1,0,-1)\ ,\eeq
and $Y^{(1,2)}$ are family-traceless. Since ${\rm Tr}(YY^{(i)})=0$, there is no appreciable kinetic mixing between the non-anomalous $U(1)$s. The $X$ charges on the three chiral families in the $\bf 27$ are of the form

\be X=(\alpha+\beta V+\gamma V')(1,1,1)\ ,\eeq
where $\alpha,~\beta,~\gamma$ are expressed in terms of the
    $X$-charges of $\overline N_i$ (=-3/2), that of ${\bf Q}\overline{\bf
    d }H_d$ (=-3), and that of the vector-like pair ,mass term
    $\overline E E$ (=-3).
 
\noindent The matter content of this model is the smallest that reproduces the 
observed quark and lepton hierarchy while cancelling the anomalies 
associated with the extra gauge symmetries:
\begin{itemize}
\item Three chiral families each with the quantum numbers of 
a $\bf 27$ of $E_6$. This means  three chiral families of the standard 
model, ${\bf Q}_i$, $\overline{\bf u}_i$, $\overline{\bf d}_i$, $L_i$, 
and $\overline e_i$, together with three right-handed neutrinos $\overline N_i$, 
three vector-like pairs denoted by $E_i$ + $\overline{\bf D}_i$ 
and $\overline E_i$ + ${\bf D}_i$, with the quantum numbers of the $\bf 5$ + 
$\overline{\bf 5}$ of $SU(5)$, and finally three real singlets $S_i$. 
\item One standard-model vector-like pair of  Higgs 
weak doublets.
\item Chiral fields that are needed to break 
the three extra $U(1)$ symmetries in the DSW vacuum. We denote these 
fields by $\theta_a$. In our minimal model with three symmetries that 
break through the FI term, we just take $a=1,2,3$. The 
$\theta$ sector  is necessarily anomalous.
\item Hidden sector gauge interactions and their matter, and other standard model singlet fields.
\end{itemize}
Finally, the charges of the three $\theta$ fields are given in terms of the matrix
\be
{\bf A}= \left( \begin{array}{ccc}
1&0&0\\ 0&-1&1\\ 1&-1&0
\end{array}  \right)\ ,\qquad
{\bf A}^{-1}= \left( \begin{array}{ccc}
1&0&0\\ 1&0&-1\\ 1&1&-1
\end{array}  \right)\ ,\ee
showing that all three fields acquire the same vacuum value. Below we display some noteworthy features of this model. 
\subsection{Quark and Charged Lepton Masses}
The Yukawa interactions in the charge $2/3$ quark sector are generated
by operators of the form
\be {\bf Q}^{}_i\bar{\bf u}^{}_jH^{}_u
{\bigl ( {\theta_1 \over M} \bigr )}^{n^{(1)}_{ij}}
{\bigl ( {\theta_2 \over M} \bigr )}^{n^{(2)}_{ij}}
{\bigl ( {\theta_3 \over M} \bigr )}^{n^{(3)}_{ij}}
\ ,\label{eq:uterm}\ee
in which the exponents must be positive integers or zero. Assuming
that only the top quark Yukawa coupling appears at tree-level, a
straighforward computation of their charges yields in the DSW vacuum
the charge $2/3$ and $-1/3$ Yukawa matrices 
 \be
Y_{}^{[u]}\sim\left( \begin{array}{ccc}
\lambda^8 &\lambda^5&\lambda^3\\ \lambda^7&\lambda^4&\lambda^2\\
\lambda^5&\lambda^2&1\end{array}  \right)\qquad,Y_{}^{[d]}\sim\lambda_{}^{3}\left( \begin{array}{ccc}
\lambda_{}^{4} &\lambda_{}^{3}&\lambda_{}^{3}\\ 
\lambda_{}^{3}&\lambda_{}^{2}&\lambda_{}^{2}\\
\lambda_{}^{}&1&1\end{array}  \right)\ .\ee
where $\lambda=\vert\theta_a\vert/ M$ is the common expansion parameter, and we have used $X^{[d]}=-3$. Diagonalization of the two Yukawa matrices yields the CKM matrix
\be
{\cal U}^{}_{CKM}\sim\left( \begin{array}{ccc}
1 &\lambda_{}^{}&\lambda_{}^{3}\\ \lambda_{}^{}&1&\lambda_{}^{2}\\
\lambda_{}^{3}&\lambda_{}^{2}&1\end{array}  \right)\ .\ee
This shows  the expansion parameter to be of the same order of magnitude 
as the Cabibbo angle $\lambda_c$. The eigenvalues of these matrices reproduce the 
geometric interfamily hierarchy for quarks of both charges
\be
{m_u\over m_t}\sim \lambda_c^8\ ,\qquad {m_c\over m_t}\sim
\lambda_c^4\ .\ee
\be
{m_d\over m_b}\sim\lambda_c^4\ ,\qquad {m_s\over m_b}\sim \lambda_c^2\
,\ee
while the quark intrafamily hierarchy is given by
\be
{m_b\over m_t}= \cot\beta\lambda_{c}^{3}\ .\ee
implying the relative suppression of the bottom to  top quark masses, 
without large $\tan\beta$.

The analysis in the charged lepton sector is similar. Since  $X^{[e]}=-3$, there are supersymmetric zeros in the $(21)$ and $(31)$ position, yielding
\be 
Y_{}^{[e]}\sim\lambda_{c}^{3}\left( \begin{array}{ccc}
\lambda_c^{4} &\lambda_c^{5}&\lambda_c^{3}\\ 
0&\lambda_c^{2}&1\\
0&\lambda_c^2&1\end{array}  \right)\ .\ee
Its diagonalization yields the lepton interfamily hierarchy
\be
{m_e\over m_\tau}\sim\lambda_c^4\ ,\qquad {m_\mu\over m_\tau}\sim
\lambda_c^2\ .\ee
Our choice of $X$ insures $X^{[d]}=X^{[e]}$, which yields  at cut-off 
\be {m_b\over m_\tau}\sim 1\ ;\ \ \sin^2\theta_w={3\over 8}~~~\leftrightarrow ~~~X^{[d]}=X^{[e]}\ .\ee
A remarkable feature of this type of model that both inter- and 
intra-family hierarchies are linked not only with one another but with 
the value of the Weinberg angle. In addition, the model predicts
a natural suppression of $m_b/m_t$, which suggests that $\tan \beta$
is of order one.

%%%%%%%%%%%%%%%%%%%%%%%%%%%%%%%%%%%%%%%%%%%%%%%%%%%%%%%%%%%%%%%%%%%%%%%%%%%

\subsection{Neutrino Masses}
Neutrino masses are naturally generated by the seesaw mechanism~\cite{SEESAW}
if the three right-handed neutrinos $\overline N_i$ acquire a Majorana mass
in the DSW vacuum. The flat direction analysis indicates that their
$X$-charges must be negative half-odd integers, with $X_{\overline N}=-3/2$. 
One finds  three massive right-handed neutrinos with masses 
\be
m_{\overline N_e}\sim M\lambda_c^{13}\ ;\qquad m_{\overline N_\mu}\sim 
m_{\overline N_\tau}\sim M\lambda_c^7\ .\ee
 In our model,
$X(L_iH_u\overline N_j)\equiv X^{[\nu]}=0$. The seesaw mechanism yields the light neutrino Yukawa matrix ($v_u \equiv\vev{H^0_u}$)
\be
{v_u^2\over M\lambda_c^3} \left( \begin{array}{ccc}
\lambda_c^6&\lambda_c^3&\lambda_c^3\\ \lambda_c^3&1&1\\ \lambda_c^3&1&1
\end{array}  \right)\ .
\label{eq:nu_matrix}\ee
A characteristic of the seesaw mechanism is that the 
charges of the $\overline N_i$ do not enter in the determination of these 
orders of magnitude as long as there are no massless right-handed
neutrinos. Hence the structure of the neutrino mass matrix depends only on
the charges of the invariants $L_iH_u$, already fixed by phenomenology and
anomaly cancellation. In particular, the family structure is determined by the lepton doublets $L_i$. In our model, since 
$L_2$ and $L_3$ have the same charges, we
 have no flavor distinction between the neutrinos of the second and third family. 
 Its diagonalization yields the neutrino mixing
matrix~\cite{MNS}
\be
 {\cal U}_{\rm MNS}=\left( \begin{array}{ccc}
1&\lambda_c^{3}&\lambda_c^{3}\\  \lambda_c^{3}&1&1
\\ \lambda_c^{3}&1&1\end{array}  \right)\ ,\ee
so that the mixing of the electron neutrino is small, of the order of
$\lambda_c^3$, while the mixing between the $\mu$ and $\tau$ neutrinos is of
order one. Remarkably enough, this mixing pattern is precisely the one
suggested by the non-adiabatic MSW ~\cite{MSW} explanation of the solar
neutrino deficit and by the oscillation interpretation of the reported
anomaly in atmospheric neutrino fluxes (which has been recently confirmed
by the Super-Kamiokande \cite{SuperK} and Soudan \cite{Soudan}
collaborations).

A naive order of magnitude diagonalization gives a
$\mu$ and $\tau$ neutrinos of comparable masses, and a much lighter electron
neutrino:
\be
m_{\nu_e}\ \sim\ m_0\, \lambda_c^{6}\ ;\qquad 
m_{\nu_\mu},\, m_{\nu_\tau}\ \sim\ m_0\ ;\qquad
m_0\ =\ {v_u^2\over M\lambda_c^3}\ ,
\label{eq:nu_mass}\ee
At first sight, this spectrum is not compatible with a simultaneous
explanation of the solar and atmospheric neutrino problems, which requires
a hierarchy between $m_{\nu_\mu}$ and $m_{\nu_\tau}$. However, the estimates
(\ref{eq:nu_mass}) are too crude: since the (2,2), (2,3) and (3,3) entries
of the mass matrix all have the same order of magnitude, the prefactors that
multiply the powers of $\lambda_c$ in (\ref{eq:nu_matrix}) can spoil the
naive determination of the mass eigenvalues. To take this effect
into account, we rewrite the neutrino mass matrix, expressed in the basis
of charged lepton mass eigenstates, as:
\be
m_0\, \left( \begin{array}{ccc}
a \lambda_c^6 & b \lambda_c^3 & c \lambda_c^3 \\ b \lambda_c^3 & d & e \\
c \lambda_c^3 & e & f
\end{array}  \right)\ ,\ee
where the prefactors $a$, $b$, $c$, $d$, $e$ and $f$, unconstrained by any
symmetry, are assumed to be of order one, say $0.5 < a, \ldots f < 2$.
Depending on their values, the two heaviest neutrinos may be either
approximately degenerate (case 1) or well separated in mass (case 2).
It is convenient  to express their mass
ratio and mixing angle in terms of the two parameters
$x = \frac{df-e^2}{(d+f)^2}$ and $y = \frac{d-f}{d+f}$:
\be
\frac{m_{\nu_2}}{m_{\nu_3}}\ =\ \frac{1-\sqrt{1-4x}}{1+\sqrt{1-4x}}\ ;
\qquad \sin^2 2 \theta_{\mu \tau}\ =\ 1\ -\ \frac{y^2}{1-4x}\ .
\label{eq:mixing_mu_tau} \ee
Case 1 corresponds to both regimes $4x \sim 1$ and $(-4x) \gg 1$, while
case 2 requires $|x| \ll 1$. Small values of $|x|$
are very generic when $d$ and $f$ have the same sign, provided that
$df \sim e^2$. Since this condition is very often satisfied by arbitrary
numbers of order one, a mass hierarchy is not less natural, given the
structure (\ref{eq:nu_matrix}), than an approximate degeneracy.

{\bf Case 1: $m_{\nu_2} \sim m_{\nu_3}$}. The
oscillation frequencies $\Delta m^2_{ij} = m^2_{\nu_j} - m^2_{\nu_i}$ are
roughly of the same order of magnitude, $\Delta m^2_{12} \sim \Delta m^2_{23}
\sim \Delta m^2_{13}$. There is no simultaneous explanation of the solar and
atmospheric neutrino data. A strong degeneracy between $\nu_2$ and $\nu_3$,
which would result in two distinct oscillation frequencies, $\Delta m^2_{23}
\ll \Delta m^2_{12} \simeq \Delta m^2_{13}$, would be difficult to achieve
unless additional symmetries are invoked. This case yields only the MSW effect, with $\Delta m^2_{12}\sim \Delta m^2_{13} \sim 10^{-6}\, eV^2$, and a total electron neutrino
oscillation probability
\be
P\, (\nu_e \rightarrow \nu_{\mu, \tau})\ =\ 4\, u^2
\lambda_c^6 \sin^2 \left( \frac{\Delta m^2_{12} L}{4 E} \right)\
+\ 4\, v^2 \lambda_c^6
\sin^2 \left( \frac{\Delta m^2_{13} L}{4 E} \right)\
\label{eq:nu_e} ,\ee
where the parameters $u$ and $v$ are defined to be $u =\frac{bf-ce}{df-e^2}$
and $v =\frac{be-cd}{df-e^2}\ $. If $\Delta m^2_{12}$ is close enough to
$\Delta m^2_{13}$, (\ref{eq:nu_e}) can be viewed as a two-flavour
oscillation with a mixing angle $\sin^2 2 \theta = 4\, (u^2+v^2)\,
\lambda_c^6$. The solar neutrino data then require $(u^2+v^2) \sim 10-20$
\cite{Langacker}, which is still reasonable in our approach. Although the
mixing between $\mu$ and $\tau$ neutrinos is of order one, they are too
light to account for the atmospheric neutrino anomaly.

{\bf Case 2: $m_{\nu_2} \ll m_{\nu_3}$}. The two distinct oscillation
frequencies $\Delta m^2_{12}$ and $\Delta m^2_{13} \simeq \Delta m^2_{23}$
can explain both the solar and atmospheric neutrino data: the non-adiabatic MSW
$\nu_e \rightarrow \nu_{\mu, \tau}$ solution suggest \cite{Langacker}
\be
 4 \times 10^{-6}\, eV^2\ \leq\ \Delta m^2\ \leq\ 10^{-5}\, eV^2
\qquad (\mbox{best fit:}\ 5 \times 10^{-6}\: eV^2)\ ,\ee
while  the atmospheric neutrino anomaly requires
\cite{atmospheric}
\be
 5 \times 10^{-4}\, eV^2\ \leq\ \Delta m^2\ \leq\ 5 \times 10^{-3}\, eV^2
\qquad (\mbox{best fit:}\ 10^{-3}\: eV^2)\ .\ee
To accommodate both, we need $0.03 \leq \frac{m_{\nu_2}}{m_{\nu_3}} \simeq x
\leq 0.15$ (with $x=0.06$ for the best fits), which can be achieved without
any fine-tuning in our model. Interestingly enough, such small values of $x$
generically push $\sin^2 2 \theta_{\mu \tau}$ towards its maximum, as can be
seen from (\ref{eq:mixing_mu_tau}). Indeed, since $d$ and $f$ have the same
sign and are both of order one, $y^2$ is naturally small compared with
$(1-4x)$. This is certainly a welcome feature, since the best fit to the
atmospheric neutrino data is obtained precisely for $\sin^2 2 \theta = 1$.

In both cases, the scale of the neutrino masses measures the cut-off
$M$. In case 1, the MSW effect requires $m_0 \sim 10^{-3}\, eV$, which
gives $M \sim  10^{18}\, GeV$. In case 2, the best
fit to the atmospheric neutrino data gives $ m_0\, (d+f) = m_{\nu_2}
+ m_{\nu_3} \simeq 0.03\, eV$, which corresponds to a slightly lower 
cut-off, $10^{16}\, GeV \leq M \leq 4 \times 10^{17}\, GeV$
(assuming $0.2 \leq d+f \leq 5$). It is remarkable that those values are
so close to the unification scale obtained by running the standard 
model gauge couplings. This result depends of course on our choice
for $X_{\overline N}=-3/2$, favored by the flat direction analysis.

To conclude this section, we note that our model predicts order-one mixing between $\nu_\mu$ and $\nu_\tau$, as well as the small angle MSW solution to the solar neutrino deficit. In addition, the scales of the measured mass eigenvalues "measure" the cut-
off to be of the order of $M_U$. Lastly,  our model predicts neither a neutrino mass in the
few $eV$ range, which could account for the hot component of the dark matter
needed to understand structure formation, nor that implied by LSND\cite{LSND}.

%%%%%%%%%%%%%%%%%%%%%%%%%%%%%%%%%%%%%%%%%%%%%%%%%%%%%%%%%%%%%%%%%%%%%%%%%
%%%%%%%%%%%%%%%%%%%%%%%%%%%%%%%%%%%%%%%%%%%%%%%%%%%%%%%%%%%%%%%%%%%%%%%%%%%%
\section{Conclusion}
The case for the extension to the standard model to an anomalous $U(1)$ is
very compelling, as it can yield the correct quark and lepton hierarchies, 
including neutrino masses and mixings in agreement by current experiments.  However, our 
model is not complete, as it only predicts
orders of magnitude of Yukawa couplings, not their prefactors.  Many of its features are found 
in free fermion theories~\cite{faraggi}, which
arise in the context of perturbative string theory. Since anomalies are involved, it is hoped that these features extend beyond perturbative string theories. Specifically,  the calculation of the Fayet-Iliopoulos term in non-perturbative regimes might rec
oncile the cut-off from the low energy theory with the string scale. 
\vskip 1cm 
{\bf Acknowledgements}
\vskip .5cm

I would like to thank Professor Nath for his kind hospitality, as well as my collaborators, J. Elwood, N. Irges and S. Lavignac, 
on whose work much of the above is based. This work was supported in part by the
United States Department of Energy under grant DE-FG02-97ER41029.

%%%%%%%%%%%%%%%%%%%%%%%%%%%%%%%%%%%%%%%%%%%%%%%%%%%%%%%%%%%%%%%%%%%%%%%%%

\end{document}